\documentclass[11pt]{article}
\usepackage{geometry} 
\usepackage{authblk}
\usepackage[numbers]{natbib}
\usepackage{graphicx}
\usepackage{amsmath,amssymb}
\usepackage{adjustbox}
\usepackage{color}
\usepackage[table]{xcolor}
\usepackage{longtable}
\usepackage[colorlinks,linkcolor=blue,citecolor=blue,urlcolor=blue]{hyperref}
\usepackage{xr}
\usepackage{array}
\providecommand{\keywords}[1]{\noindent\textbf{Keywords --} #1}
\newcommand{\figsdirsecond}{./}
\newcommand{\beginsupplement}{
        \setcounter{table}{0}
        \renewcommand{\thetable}{S\arabic{table}}%
        \setcounter{figure}{0}
        \renewcommand{\thefigure}{S\arabic{figure}}% 
        \setcounter{equation}{0}
        \renewcommand{\theequation}{S\arabic{equation}}
        \setcounter{section}{0}
     }
\newcommand{\new}[1]{\textcolor{black}{#1}}

\title{\bf Metabolic basis of brain-like electrical signalling in bacterial 
communities}
\author[1]{\bf Rosa Martinez-Corral}
\author[2,*]{\bf Jintao Liu}
\author[3,*]{\bf Arthur Prindle}
\author[4]{\bf G\"urol M. S\"uel}
\author[1]{\bf Jordi Garcia-Ojalvo}
\affil[1]{\small\it Department of Experimental and Health Sciences, Universitat 
Pompeu Fabra,
Barcelona Biomedical Research Park, 08003 Barcelona, Spain}
\affil[2]{\small\it Center for Infectious Diseases Research and Tsinghua-Peking 
Center for Life Sciences, School of Medicine, Tsinghua University, 100084 
Beijing, China}
\affil[3]{\small\it Department of Biochemistry and Molecular Genetics, Feinberg 
School of Medicine, Northwestern University, Chicago, IL 60611, USA; Center for 
Synthetic Biology, Northwestern University, Evanston, IL 60208, USA.}
\affil[4]{\small\it Division of Biological Sciences, San Diego Center for 
Systems Biology, and Center for Microbiome Innovation, University of California 
San Diego, California 92093, USA}
\affil[*]{Equal contribution}
\date{} 

\begin{document}

\maketitle


\begin{abstract}
Information processing in the mammalian brain relies on a careful regulation of the membrane potential dynamics of its constituent neurons, which propagates across the neuronal tissue via electrical signalling.
We recently reported the existence of electrical signalling in a much simpler organism, the bacterium \emph{Bacillus subtilis}.
In dense bacterial communities known as biofilms, nutrient-deprived {\em B. subtilis} cells in the interior of the colony use electrical communication to 
transmit stress signals to the periphery, which interfere with the growth of 
peripheral cells and reduce nutrient consumption, thereby relieving stress from 
the interior.
Here we explicitly address the interplay between metabolism and 
electrophysiology in bacterial biofilms, by introducing a spatially-extended 
mathematical model that combines the metabolic and electrical components of the 
phenomenon in a discretised reaction-diffusion scheme.
The model is experimentally validated by environmental and genetic 
perturbations, and confirms that metabolic stress is transmitted through 
the bacterial population via a potassium wave.
Interestingly, this behaviour is reminiscent of cortical spreading depression in 
the brain,  characterised by a wave of electrical activity mediated by potassium 
diffusion that has been linked to various neurological disorders, calling for future studies on the
evolutionary link between the two phenomena.
\end{abstract}

\section{Introduction}

Local interactions between individual cells are known to generate complex emergent behaviour in multicellular organisms, which underlie the functionalities of tissues, organs, and physiological systems \citep{Murray:2001si}.
The human brain provides one of the finest examples of such phenomena, with its 
complexity ultimately emerging from the interactions between neuronal cells \cite{Freeman:2000wt}. 
Similarly, unicellular organisms can also self-organise into communities with 
complex community-level phenotypes \cite{Maree:2001bq}. \new{Bacterial biofilms, in particular,} provide a good model system for the study of collective 
behaviour in biological systems \citep{Flemming2016}.

\begin{figure}[htbp]
{\includegraphics[width=\textwidth]{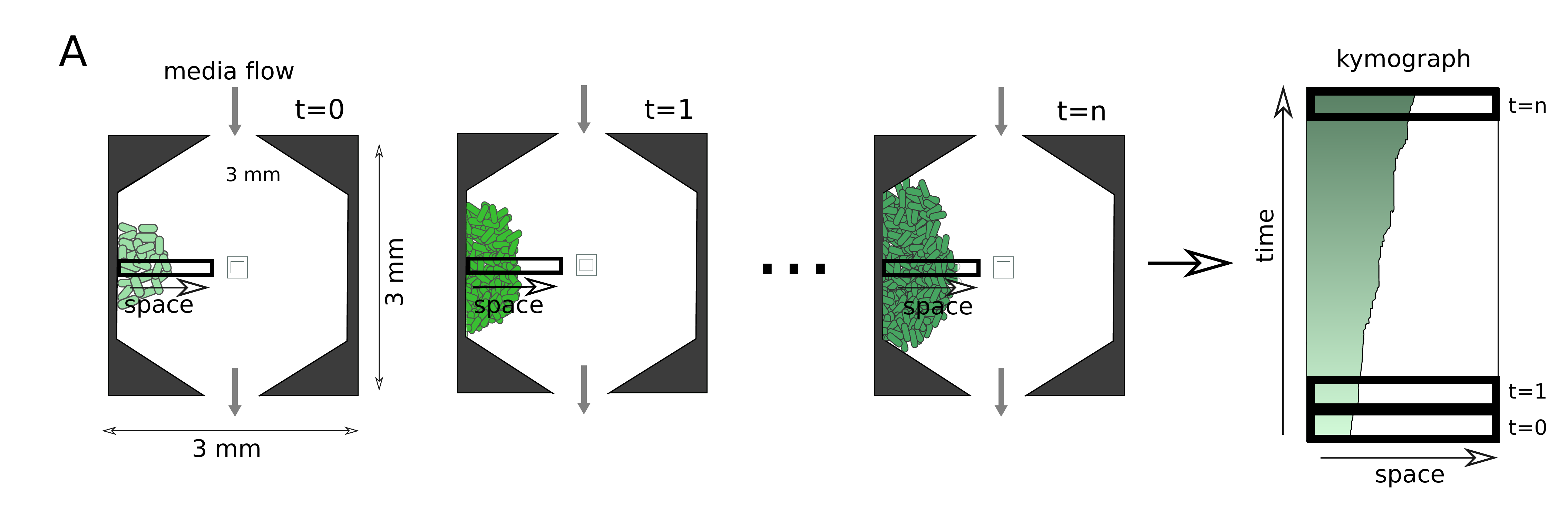}}
{\includegraphics[width=\textwidth]{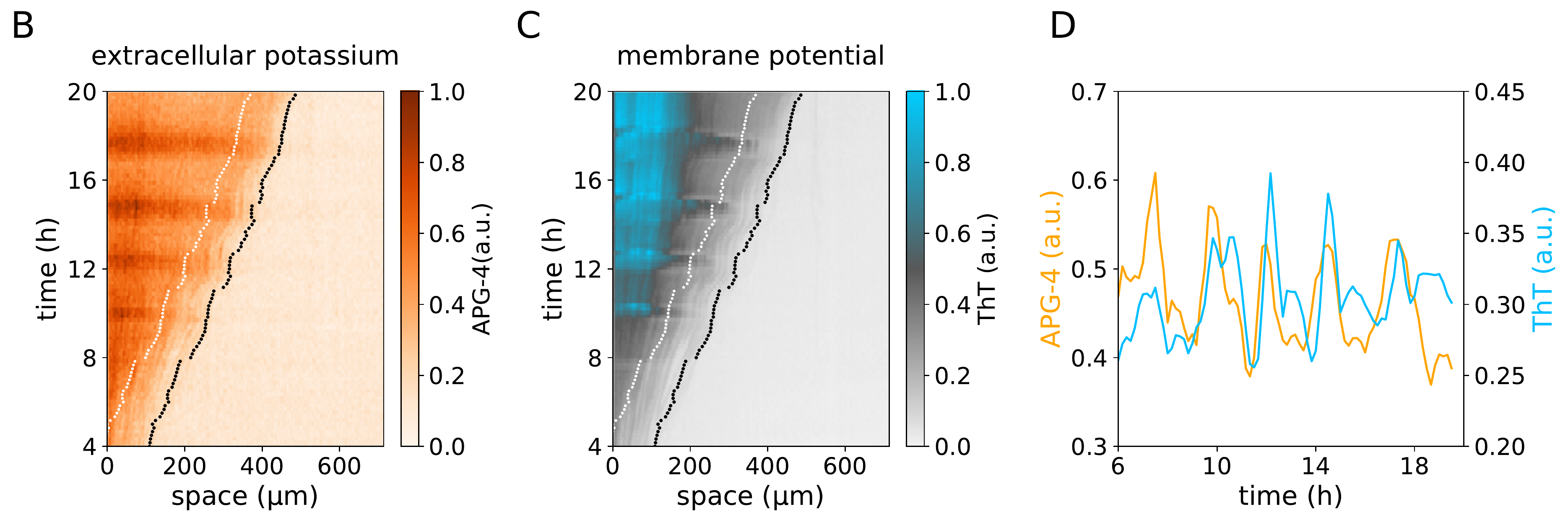}}
\caption{Biofilms of \emph{B. subtilis} display oscillations in growth rate and 
electrical signalling activity. A)~Scheme of the microfluidics device used to 
grow biofilms. Biofilms grow around the central pillar or attached to the wall, 
as in the cartoon. Biofilm dynamics can be represented as a kymograph, where 
each row represents a one dimensional cross-section of the biofilm area at a 
given movie frame. B-D)~Experimental kymographs of a cross-section of a biofilm, 
as depicted in A, for different channels. Each pixel is the average of a region 
of $5.2 \times 5.2$ $\mu m$. The dotted black line denotes the limit of the 
biofilm as established from the phase-contrast images. B)~Oscillations in 
extracellular potassium, reported by APG-4. C)~Oscillations in membrane 
potential, reported by ThT.   D)~Extracellular potassium and ThT time traces 
corresponding to the position denoted by the white line in panels B,C. Data has 
been smoothed with an overlapping sliding window of size 3 frames.  
}\label{fig_1}
\end{figure}

We have previously shown that biofilms of the bacterium \emph{B. subtilis} can 
display community-level oscillatory dynamics \citep{Liu2015}. These oscillations 
ensure the viability, under low-nitrogen (low-glutamate) conditions, of the 
cells in the centre of the community, which are crucial for community regrowth 
upon external sources of stress like antibiotics. 
In our experimental setup (Fig.~\ref{fig_1}A), growing two-dimensional biofilms 
that reach a certain size develop oscillations \citep{Martinez-Corral2018}, such 
that peripheral cells periodically stop their growth (see black dotted line in 
Fig.~\ref{fig_1}B,C). This behaviour was attributed to a metabolic codependence 
between central and peripheral cells mediated by ammonium \cite{Liu2015} 
(see introduction to Section \ref{sec_full} below). 

\begin{figure}[htb]
\centerline{\includegraphics[width=0.75\textwidth]{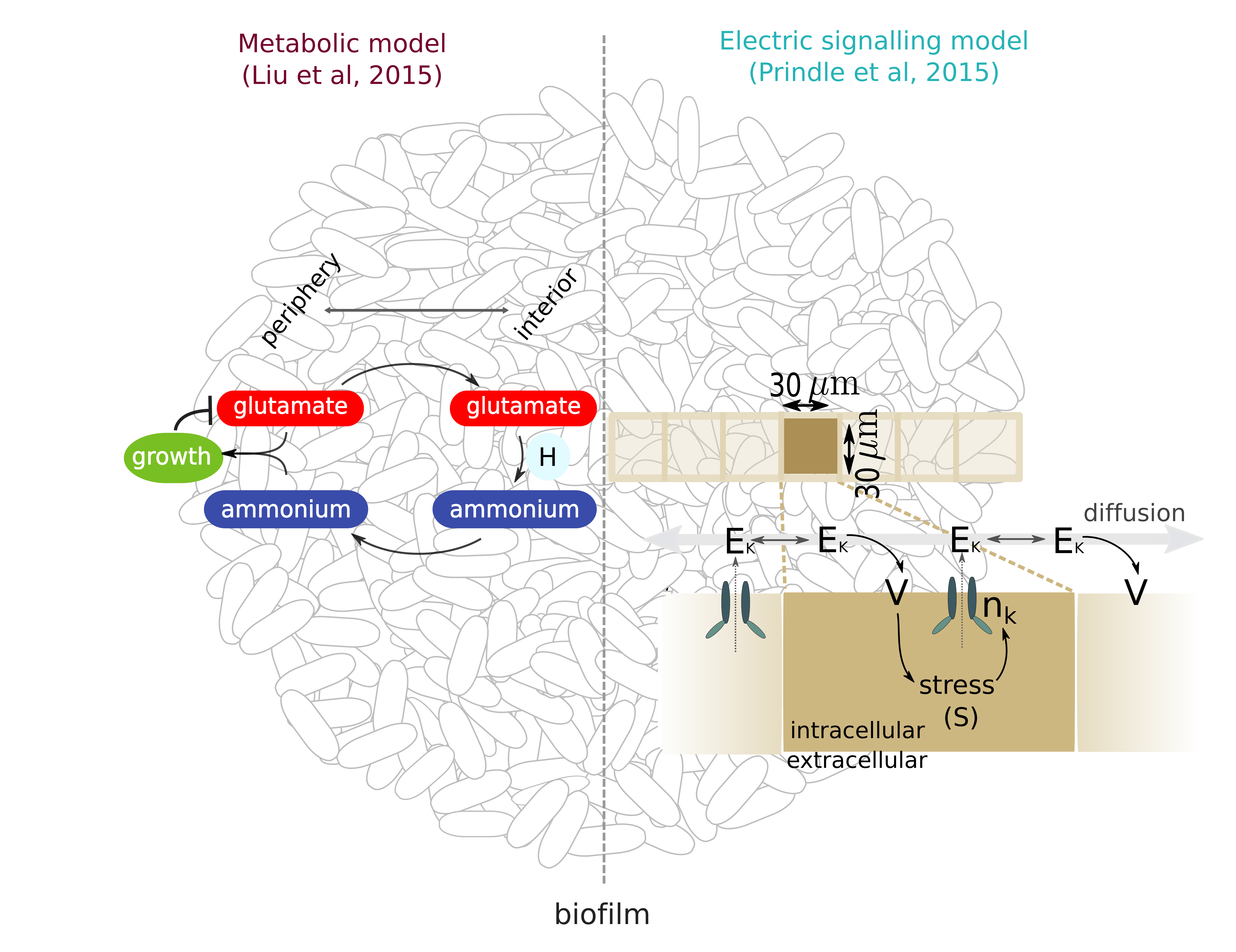}}
\caption{Metabolic (left) and electric (right) processes involved in the 
response of the biofilm to a gradient of nutrient-limitation-drive stress from 
the interior to the periphery. The left-hand side schematises the interactions 
considered in the metabolic model of Ref. \citep{Liu2015}, whereas the 
right-hand side schematises the electrical signalling propagation model proposed 
in Ref. \citep{Prindle2015}.
In the diagram on the left, H represents the enzyme GDH.
In the diagram on the right, E$_{\rm K}$ stands for excess extracellular 
potassium, V for the membrane potential, and n$_{\rm k}$ for the gating variable 
of the potassium channel.}
\label{fig_2}
\end{figure}

Besides exhibiting metabolic oscillations in growth rate, our biofilms display 
periodic changes in extracellular potassium (Fig.~\ref{fig_1}B) and cellular 
membrane potential (Fig.~\ref{fig_1}C), as reported by the fluorescent dyes 
APG-4 and Thioflavin-T (ThT), respectively.
The membrane potential changes reported by ThT are caused by the periodic 
release of intracellular potassium ions (Ref. \citep{Prindle2015} and 
Fig.~\ref{fig_1}D).
Potassium propagates from the centre of the biofilm to its periphery according 
to the following scenario: potassium released by glutamate-deprived (and thus 
metabolically stressed) cells in the centre diffuses towards 
neighbouring cells, causing depolarisation, stress and subsequent potassium 
release (Fig.~\ref{fig_2}-right), which thereby keeps propagating along the 
biofilm in a bucket-brigade manner \citep{Prindle2015}.

Notably, \new{potassium and glutamate are the two most abundant ions in living cells \cite{Gundlach2018}. In addition to their their roles in osmoregulation, pH maintenance and nitrogen metabolism, they are essential for information propagation in animal nervous systems. In particular,} potassium is one of the key ions involved in the modulation of the 
membrane potential in neurons, and glutamate is a central excitatory 
neurotransmitter. \new{Besides the presence of the same ions in the two scenarios, there are also phenomenological links.}  In a normally functioning brain, electrical signalling typically propagates 
along neurons through action potentials mediated by sodium and potassium that can travel at speeds up to meters 
per second \citep{Hall2011}. In addition, \new{under pathological conditions}, electrical activity can also propagate over the brain on a much slower timescale, on the order of 
millimetres per minutes (reviewed in Ref. \citep{Ayata2015}), similar to the biofilm potassium waves.
This occurs during the phenomenon of cortical spreading depression, a 
wave of intense depolarisation followed by inhibition of electrical activity that entails a strong metabolic disturbance of neuronal function.
Cortical spreading depression has been shown to occur in the brains of multiple 
species, from insects \citep{Rodgers2007, Spong2016} to mammals 
\citep{Lauritzen2011}, \new{where it has been regarded as an evolutionary conserved process \citep{Ayata2015}} that has been linked to human pathologies like migraine, 
and to the propagation of brain damage during ischemic stroke.
The phenomenon has been modelled as a reaction-diffusion process 
\citep{Dahlem2004, Chang2013} according to the following mechanism: an initial 
stimulus leads to an increase in extracellular potassium, causing the membrane 
potential to rise beyond a critical threshold that triggers a self-amplifying 
release of potassium, a major ionic redistribution of sodium, calcium and 
chloride, and release of neurotransmitters like glutamate.
Diffusion of potassium and glutamate to neighbouring cells leads to subsequent 
depolarisation beyond a threshold, thus causing ionic release and 
self-propagation of the wave \citep{Ayata2015,Dahlem2003, Enger2015}.

The \new{widespread importance of potassium and glutamate, and the links between electrical signalling in bacteria and neuronal dynamics} led us to investigate in more detail the roles of those two ions in the oscillations exhibited by \emph{B. subtilis}.
In this work, we unify in a discretised reaction-diffusion scheme the metabolic 
and electrical components of the system.
Assuming a homogeneous population of cells, the model reveals the spontaneous 
emergence of two phenotypically-distinct populations of cells as a consequence 
of the spatiotemporal dynamics of the system, which causes oscillations to start 
beyond a critical size. \new{We begin by developing a model that accommodates the interactions considered in previous works both for the metabolic and electrical components of the phenomenon. Then we show that the details of the ammonium metabolism are not required to explain the key aspects of the dynamics. T}he behaviour can be explained by a model with minimal interactions between 
glutamate metabolism and potassium signalling, thus clarifying the process in 
bacteria and providing further insight into the roles of these ubiquitous 
biological ions. 

\section{A unified spatially-extended model of biofilm oscillations}
\label{sec_full}

Conceptually, biofilm oscillations can be understood to arise as result of a delayed, spatially-extended negative feedback of metabolic stress \citep{Martinez-Corral2018}.
A first molecular model was proposed in Ref.~\citep{Liu2015} 
(left half of Fig.~\ref{fig_2}),  based on the fact that cells need to create glutamine 
from glutamate and ammonium for biomass production.
Since ammonium is not present in the MSgg medium used in the experiments, it 
must be synthesised by the cells from glutamate.
The model assumed that only cells in the interior of the biofilm produce 
ammonium (through the glutamate dehydrogenase enzyme GDH, \new{H in 
Fig.~\ref{fig_2}}), part of which diffuses away to the periphery of the 
biofilm, where it is used by the peripheral cells to grow.
In turn, peripheral growth reduces glutamate availability in the centre, and 
subsequently ammonium availability in the periphery.
As a result, peripheral cells stop growing, thus allowing the centre to recover 
glutamate and produce ammonium again, leading to a new oscillation cycle.
The model considered two different cell populations --interior and peripheral--, 
and was subsequently implemented on a continuous spatial domain in Ref.~\cite{Bocci2018}, with still pre-defined interior and peripheral cell types. 

\new{This model can readily generate oscillations, which crucially rely 
on the assumption of a strongly nonlinear activation of GDH by glutamate in the interior 
population of cells (modelled with a Hill function with large Hill coefficient). However, it does not account for the electrical 
signalling component of the phenomenon, which we incorporate next.}

\subsection{Full model description}
\begin{figure}[htbp]
\centerline{\includegraphics[width=0.75\textwidth]{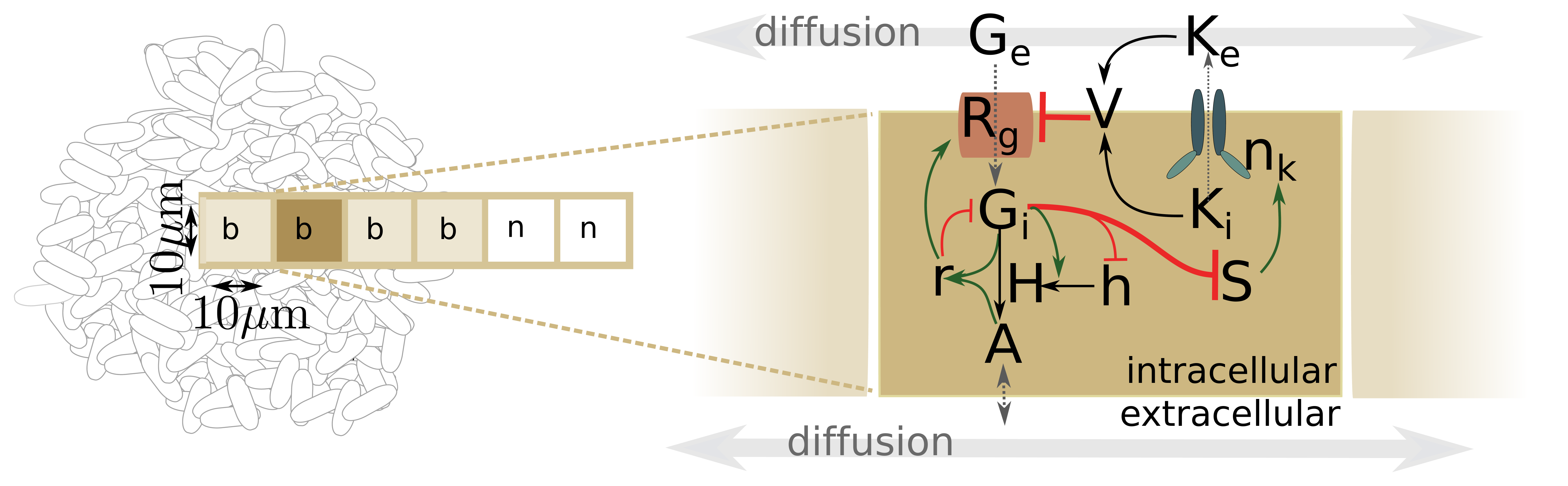}}
\caption{Scheme of the unified metabolic-electric model introduced in this work.
In the lattice scheme on the left, the labels 'b' and 'n' inside the sites
denote 'biofilm' and 'non-biofilm' lattice sites, respectively.
In the network diagram on the right, G, K, h, H and A represent glutamate, 
potassium, inactive GDH, active GDH and ammonium, respectively.
The quantity r represents biomass-producing biomolecules such as ribosomal 
proteins.
R$_{\rm g}$ denotes the glutamate receptor, V the membrane potential, $n_{\rm k}$ the gating variable of the potassium channel, and S the stress.
h stands for inactive GDH.
The subindices 'i' and 'e' label intracellular and extracellular quantities, 
respectively. The two thick flat arrows highlight the links between the metabolic and electrical components of the system.}
\label{fig_3_0}
%%\end{adjustwidt/Users/rosamartinezcorral/Dropbox (HMS)/revisionb/latex/supplementaryv1.pdfh}
\end{figure}

We consider a horizontal cross-section through the middle of a biofilm (as 
depicted in Fig.~\ref{fig_1}A) which allows us to simplify the system into a 
one-dimensional lattice, as shown in Fig.~\ref{fig_3_0}.
The left-most lattice sites are `biofilm' (shaded squares, 'b'), followed by 
\new{`non-biofilm'} sites (white squares, 'n') on the right.
\new{We begin by a more detailed model (Fig.~\ref{fig_3_0}) that includes all the aforementioned metabolic 
interactions \citep{Liu2015} (left half of Fig.~\ref{fig_2}) in addition to the 
electrical signalling propagation \citep{Prindle2015} (right half of Fig.~\ref{fig_2}).
The details of the equations can be found in the Supplementary, and we next describe the main points of the model (see also Table~\ref{table_ass})}.

\new{`Non-biofilm' sites have extracellular variables only: extracellular glutamate ($G_e$), ammonium ($A$, assumed to have the same concentration inside and outside cells, following \citep{Liu2015}), and extracellular potassium $K_e$. These variables diffuse between neighboring lattice sites. In addition, we also account for the fact that media is flowing constantly through the microfluidics chamber such that their concentration tends to become close to that in the media (Supplementary Eqs.~(\ref{eq:Ge}), (\ref{eq:A}), (\ref{eq:Ke})). }

\new{`Biofilm' sites correspond to small biofilm regions, such that variable values would correspond to averages across multiple cells. As schematised in Fig.~\ref{fig_3_0}, in this sites there are also the extracellular variables (also subject to diffusion and media flow) in addition to the intracellular ones that take part in the various biochemical reactions. We model explicitly the dynamics of extracellular ($G_e$) and intracellular 
($G_i$) glutamate in the biofilm (Supplementary Eqs.~(\ref{eq_Ge})-(\ref{eq_Gi})), where we consider glutamate uptake into the cells and the conversion of glutamate into ammonium by the action of the enzyme GDH. In turn, we model GDH synthesis in its inactive form ($h$) and subsequent activation ($H$) (Supplementary Eqs.~(\ref{eq_h})-(\ref{eq_H})). We assume that high concentrations of glutamate inhibit GDH synthesis. This follows from the original observation that GDH overexpression in the periphery kills biofilm oscillations \citep{Liu2015}, suggesting that GDH expression is restricted to the centre, where glutamate levels are low. Moreover, we assume that there is a threshold for GDH activation by glutamate, as in the original metabolic model. Glutamate and ammonium are assumed to be used for the production of biomass-producing biomolecules such as ribosomal proteins, denoted by $r$ (Supplementary Eq.~(\ref{eq_r})). We also account explicitly for the glutamate transporter concentration ($R_{g}$), 
which we assume to saturate for large enough $G_e$. In order to accommodate the assumption that cells with a higher metabolic 
activity consume more glutamate, we consider that $R_{g}$ is subject to an 
inducible activation that depends on the presence of biomass-producing 
biomolecules (Supplementary Eq.~(\ref{eq_Rg})). }

\new{In \citep{Prindle2015}, the YugO potassium channel, which has a TrkA gating domain, was determined to be responsible for potassium release in this phenomenon. TrkA proteins are potassium channel regulatory proteins with nucleotide-binding domains that bind NAD/NADH \cite{Schlosser1993, Roosild2002} and ADP/ATP \cite{Cao2013}, thus coupling potassium translocation to the metabolic state of the cell. Glutamate is a central amino acid at the cross-road between carbon and nitrogen metabolism. Our assumption is that low glutamate levels lead to imbalances in the aforementioned nucleotides that are sensed by TrkA and trigger potassium release \cite{Prindle2015}. Therefore, we assume that high levels of intracellular glutamate inhibit the production of stress-related biomolecules, resulting in the following equation for stress dynamics:}
\begin{align}
\frac{dS}{dt}&=\frac{S_0}{1+\bigl(\frac{G_i}{G_{s0}}\bigr)^{n_s}}-\gamma_s\,
S\label{eq_S}
\end{align} 
\new{In turn, $S$ enhances the opening probability of the potassium channel, given by $n_k$.
This results in a link from the metabolic component of the phenomenon to the electrical counterpart.}

\new{As the potassium channel opens, potassium is released to the extracellular environment. Conversely, potassium uptake is assumed to be governed by homeostatic processes that tend to 
keep its intracellular concentration at a fixed value, in addition to depend on the cellular metabolic state (glutamate level), to account for the energy demand of the process (Supplementary Eqs.~(\ref{eq_Ke})-(\ref{eq:Ki})). 
Extracellular and intracellular potassium concentrations determine the membrane potential ($V$), whose dynamics is described by a Hodgkin-Huxley-like conductance-based model (Supplementary Eq.~(\ref{HHeq})).
Finally we include the ThT reporter $\mathcal{T}$ downstream of the membrane 
potential, increasing when the cells become hyperpolarised due to potassium 
release (Supplementary Eq.~(\ref{eq_T})).}

In \emph{B. subtilis}, glutamate is imported by pumps such as the symporter GltP 
\citep{Kunst1997}, which uses the proton motive force as a source of energy, and 
this is in turn influenced by the membrane potential \citep{Tolner1995}.
We therefore consider that glutamate transport into the cell is modulated by the 
membrane potential $V$, such that depolarisation reduces entry, and 
hyperpolarisation enhances it.
This provides the link from the electrical to the metabolic component of the phenomenon, and is represented with the following switch-like import-modulation term:
\begin{align}
{\cal F}(V)&=\frac{1}{1+\exp(g_v\,(V-V_0))}\label{eq_Ginip}
\end{align}

The biofilm is assumed to expand proportionally to the biomass-producing biomolecules. 
In order to simulate growth, we consider that `non-biofilm' lattice sites neighbouring 
biofilm sites become occupied with cells (and thus become a lattice site of type 
`biofilm') with probability $P_{grow}\,dt\,r$, with $dt$ being the simulation 
time step.
Importantly, each new biofilm lattice site inherits the intracellular variables 
of the `mother' site. Due to this particularity, the continuum approximation is 
unsuitable, and we simulate the system as a coupled map lattice.

\begin{center}
{\rowcolors{2}{gray!15}{white}
\begin{table}
\linespread{0.85}\selectfont
\renewcommand{\arraystretch}{1.5}
\begin{tabular}{>{\raggedright}m{8cm}|>{\raggedright}m{3cm}|>{\raggedright\arraybackslash}m{3.25cm}}       
\hline
{\bf Model assumption}               & {\bf Full model (section \ref{sec_full})}     & {\bf Simplified model (section \ref{sec_electricG})} \\\hline
Extracellular glutamate flow and diffusion                                                                       & Yes                                                   & Yes                                            \\
Ammonium flow and diffusion                                                                                    & Yes                                                   & No (no ammonium present)                       \\
Extracellular potassium flow and diffusion                                                                 & Yes                                                   & Yes                                            \\
Decay of flow and diffusion rates towards the biofilm interior as a consequence of cellular density            & Yes                                                   & Yes                                            \\
Glutamate uptake favoured by hyperpolarisation and disfavoured by depolarisation                               & Yes                                                   & Yes                                            \\
Glutamate uptake enhanced in metabolically active cells                                                  & Through the effect of biomass-producing biomolecules & Directly linked to intracellular glutamate   \\
Saturable glutamate uptake                                                                                     & Yes                                                   & Yes                                            \\
GDH dynamics regulated by glutamate                                                                        & Yes                                                   & No (no GDH present)                            \\
Ammonium produced from intracellular glutamate by GDH                                                         & Yes                                                   & No (no ammonium present)                       \\
Growth favored by high levels of intracellular glutamate                                                        & Through intermediate biomass-producing biomolecules.  & Directly modulated by intracellular glutamate \\
Metabolic stress inhibited by intracellular glutamate                                                         & Yes                                                   & Yes                                            \\
Potassium channel opening probability ($n_k$) increased by stress                                              & Yes                                                   & Yes                                            \\
Potassium released as a result of the opening probability ($n_k$) and membrane potential                          & Yes                                                   & Yes                                            \\
Potassium uptake governed by homeostatic processes and the metabolic state of the cells (intracellular glutamate) & Yes                                                   & Yes                                            \\
Membrane potential determined by intracellular and extracellular potassium concentrations                       & Yes                                                   & Yes                                            \\
ThT as a reporter of membrane potential                                                                  & Yes                                                   & Yes                                           
\end{tabular}
\caption{\new{Main assumptions and comparison between the full and simplified models.}}
\label{table_ass}
\end{table}
}
\end{center}

\subsection{Simulations of the model}

The diffusion coefficient in water for ions and small molecules such as 
potassium and glutamate is of the order of $\sim 10^{6}\mathrm{\mu m}^2$/h.
We thus fixed the diffusion coefficient of potassium and ammonium in the media 
to this value, and assumed the  diffusion coefficient of glutamate to be half 
that value, due to its larger molecular weight.
Regular glutamate concentration in the media is 30~mM and potassium 
concentration is 8~mM. Since there is no ammonium in the medium, $A_m=0$. We 
fixed the resting membrane potential to $-150$~mV. Furthermore, oscillations 
experimentally start at a biofilm size of around 600 $\mathrm{\mu m}$ with a 
period of around 2~hours \citep{Martinez-Corral2018}, and potassium and ThT 
signals should be correlated \citep{Prindle2015}. With these constraints, we 
manually adjusted the rest of the parameters (Table \ref{table_extended}) in 
order to reproduce the experimentally observed oscillatory dynamics. 

\begin{figure}[htbp]
\centerline{\includegraphics[width=\textwidth]{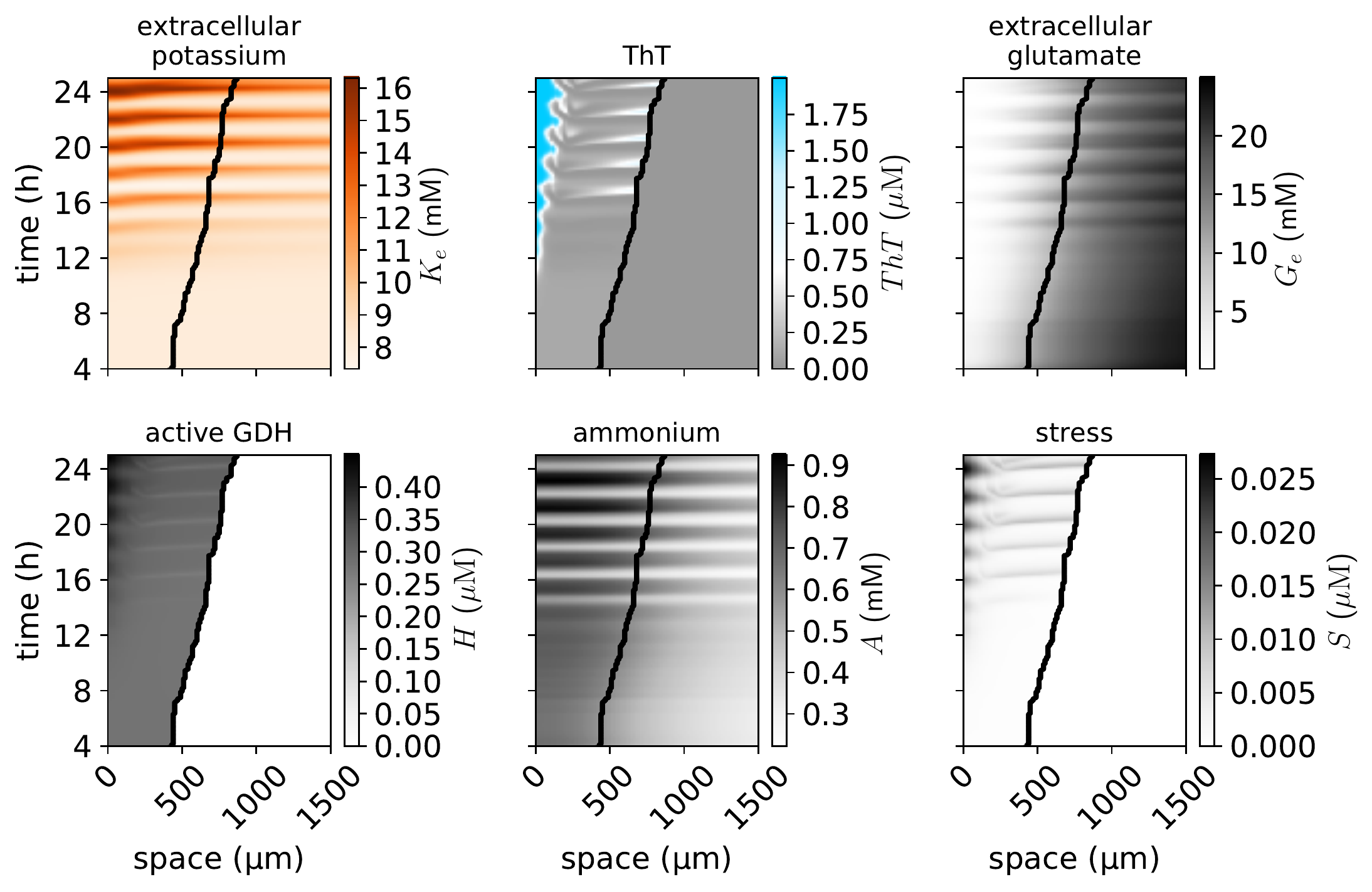}}
\caption{Oscillations in the full metabolic-electrophysiological model.
Kymographs of the model simulation results, showing oscillations in various 
model variables.} \label{fig_3}
%%\end{adjustwidth}
\end{figure}

Figure~\ref{fig_3} shows kymographs for extracellular potassium and ThT (variables that can be 
experimentally monitored as explained above, see Fig.~\ref{fig_1}), as well as 
for extracellular glutamate, stress, active GDH and ammonium, which have not 
been experimentally quantified. Oscillations with a period of about two hours 
emerge in all variables once the biofilm becomes large enough. Moreover, in 
agreement with the experimental data, the top left plot shows that potassium 
peaks are likely to coincide with periods of no growth. 

In contrast to the original model of Ref.~\citep{Liu2015} (see also 
Ref.~\citep{Bocci2018}), where two populations of cells (central and peripheral) 
were pre-defined regarding GDH production, in the current model an \emph{a 
priori} separation of cell types is no longer required, but emerges 
spontaneously.
This can be seen in the bottom left kymograph of Fig.~\ref{fig_3}, which shows 
that a central population with higher levels of GDH activity and stress emerges 
over time.
Notably, this model does not require strong cooperativity in GDH activation for 
the oscillations to emerge ($n_{H}$ = 2, in contrast to the value of 7 in Ref. 
\citep{Liu2015} and 12 in Ref. \citep{Bocci2018}).
This is likely to be the result of the electrical component of the model.
In order to test the relevance of this component of the oscillations, we next 
simplify the metabolic details and consider only the interplay between glutamate 
metabolism and electrical signalling.

\section{Oscillations emerge from the interplay between glutamate and electrical 
signalling}\label{sec_electricG}

\begin{figure}[htb!]
\adjustbox{valign=t}{\begin{minipage}{0.63\textwidth}
\vspace{2mm}
\includegraphics[width=\textwidth]{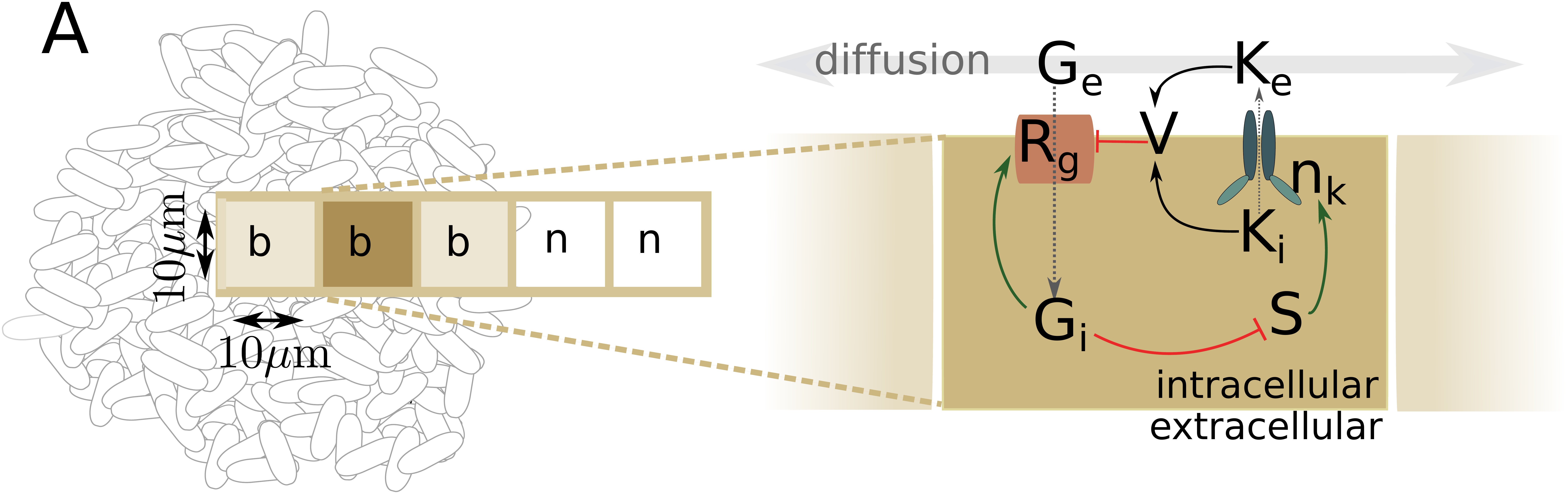}
\end{minipage}}
\adjustbox{valign=t}{\begin{minipage}{0.34\textwidth}
{\includegraphics[width=\textwidth]{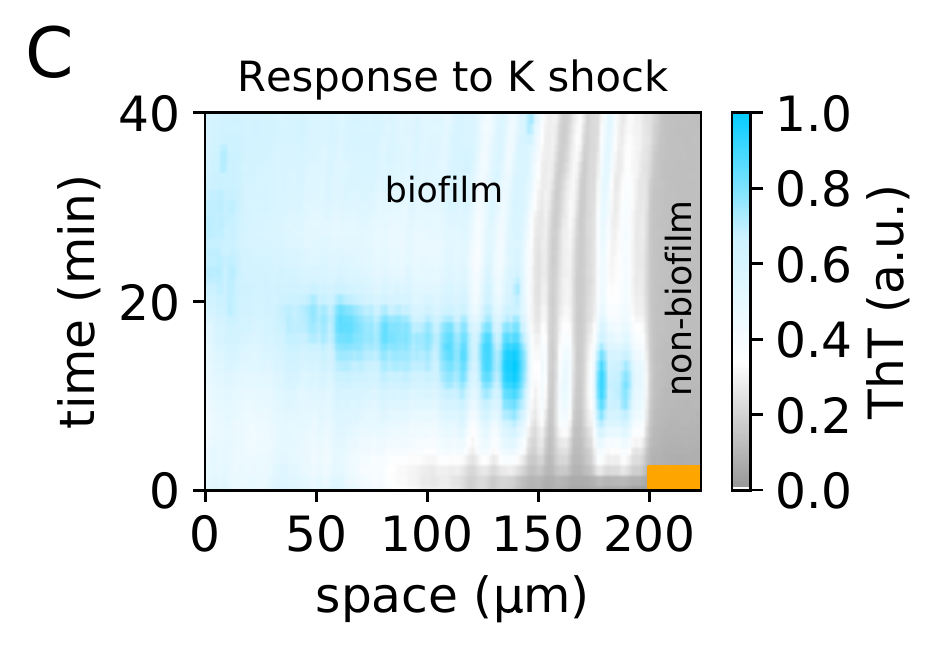}}
\end{minipage}}
{\includegraphics[width=\textwidth]{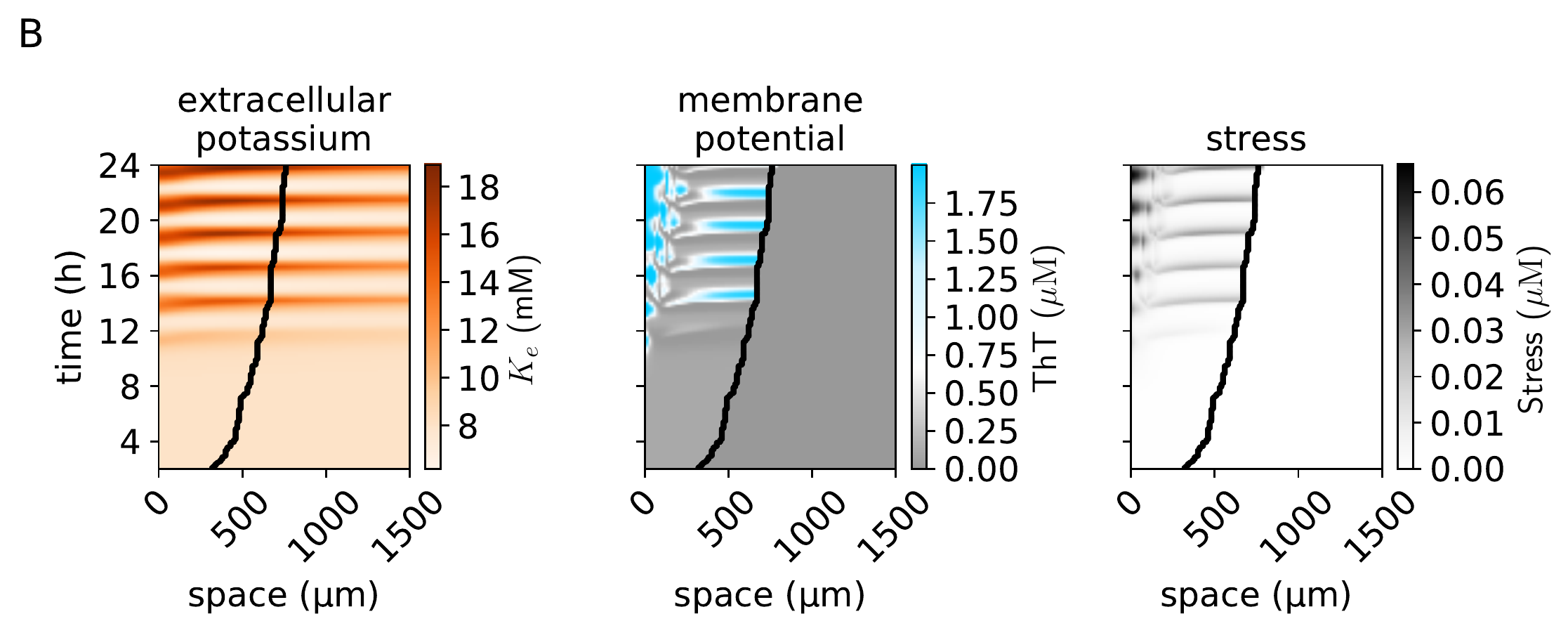}}
\caption{Oscillations in a simplified model, with glutamate and the electrical 
signalling components. A)~Scheme of the model. B)~Kymographs of various 
variables of the system. C) Electrical signalling wave (reported by ThT) in an 
experimental biofilm triggered by an increase of potassium in the media from 8 
mM to 300 mM during 3 minutes (denoted by the orange mark). The wave propagates 
inwards, from the periphery (right) to the centre (left). }\label{fig_electric}
\end{figure}

We now simplify the metabolic aspects of the full model introduced 
above, eliminating GDH, ammonium and biomass-producing biomolecules  (Supplementary Eqs.~(\ref{eq_h})-(\ref{eq_r})), and supposing 
instead that intracellular glutamate directly increases the glutamate 
transporter and directs growth (Fig.~\ref{fig_electric}A). With this, the equations governing the metabolic part of the phenomenon become:
\begin{align}
\frac{dG_e}{dt}&=-\alpha_{g}\,{\cal 
F}(V)\,R_{g}\,\frac{G_e}{k_{g}+G_e}+\Lambda_{\phi}\,\phi\,(G_{m}-G_e)+\Lambda_{D
}\, D_g\nabla^2\,G_e \label{eq_Gem}\\
\frac{dG_i}{dt}&=\alpha_{g}\,{\cal 
F}(V)\,R_{g}\,\frac{G_e}{k_{g}+G_e}-\delta_g\,G_i\\
\frac{dR_{g}}{dt}&=\alpha_{R}-\delta_{R}\,R_g+\alpha_{r}\frac{G_i^{n_r}}{k_{r}^{
n_r}+G_i^{n_r}}
\end{align}
in addition to considering glutamate diffusion in the `non-biofilm' sites (Supplementary Eq.~\ref{eq:Ge}).

In this case, the probability with which the biofilm grows depends 
directly on the concentration of intracellular glutamate at the biofilm boundary 
according to $P_{grow}\,dt\,G_i$. 
The dynamics of the electrical 
part of the model remain unchanged (Supplementary Eqs.~(\ref{eq_SA})-(\ref{eq_T})).

The kymographs in Fig.~\ref{fig_electric}B show that oscillations beyond a 
critical size also emerge in this case. The results suggest the following 
mechanism for the oscillations: once the biofilm reaches a critical size, 
glutamate in the centre of the biofilm is too low and leads to metabolic stress. 
As a result, cells release potassium, that actively propagates towards the edge 
due to depolarisation and subsequent metabolic stress. 
Depolarisation leads to a drop in glutamate consumption levels in the peripheral 
region, due to the immediate effect on the transport efficiency, and a slightly 
delayed effect through the reduction in the glutamate transporter. As a result, 
glutamate can diffuse inwards to allow stress relief in the centre, and a new 
oscillation cycle can start. 

\begin{figure}
\centerline{\includegraphics[width=0.7\textwidth]{\figsdirsecond/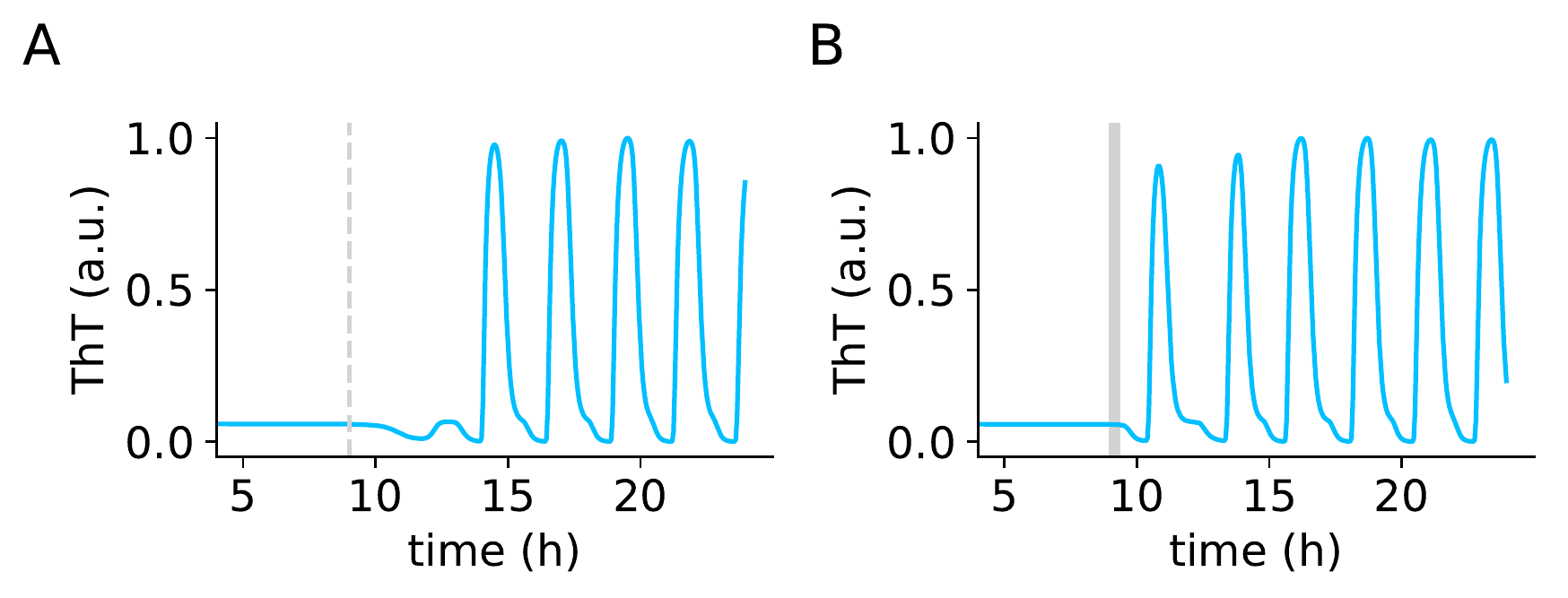}}
\caption{Stop-flow triggers oscillations in the simplified model. Normalised ThT 
time traces at the periphery (50 $\mathrm{\mu m}$ from the biofilm edge). 
A)~Reference simulation. Vertical dashed line indicates the time of stop-flow in 
B. (This simulation is the same as in Fig.~\ref{fig_electric}B). B)~The biofilm 
was perturbed with a stop-flow-like perturbation ($\phi=0$ during 20 min at 
$t=9$ hours). The growth noise realisation is the same in both simulations, such 
that the only difference is the perturbation. }
\label{fig_stopflow}
\end{figure}

According to this, oscillation onset requires stress in the centre to surpass a 
threshold that leads to potassium release. As we have recently described 
\citep{Martinez-Corral2018}, oscillations can be triggered in experimental 
biofilms by transiently stopping the media flow of the microfluidics chamber. 
This is also the case in this model, where oscillations can be triggered by 
transiently setting the flow-rate parameter ($\phi$) to zero. If the biofilm is 
sufficiently large, such a perturbation leads to a sudden reduction in the 
glutamate availability and stress increase, with the subsequent potassium 
release and oscillatory onset (Fig.~\ref{fig_stopflow}). 

Moreover, the model predicts that depolarising any region of the biofilm should 
be sufficient to trigger a self-propagating wave of stress and potassium 
release, with the associated changes in membrane potential.
This is reproduced experimentally as shown in Fig.~\ref{fig_electric}C: a short 
increase in the concentration of potassium in the media triggers a wave in 
experimental biofilms, which propagates inwards from the original (peripheral) 
depolarisation site.

\section{Glutamate metabolism and stress release determine oscillation onset 
size and period}
\label{sec_onset}

\begin{figure}[htb!]
\centerline{\includegraphics[width=\textwidth]{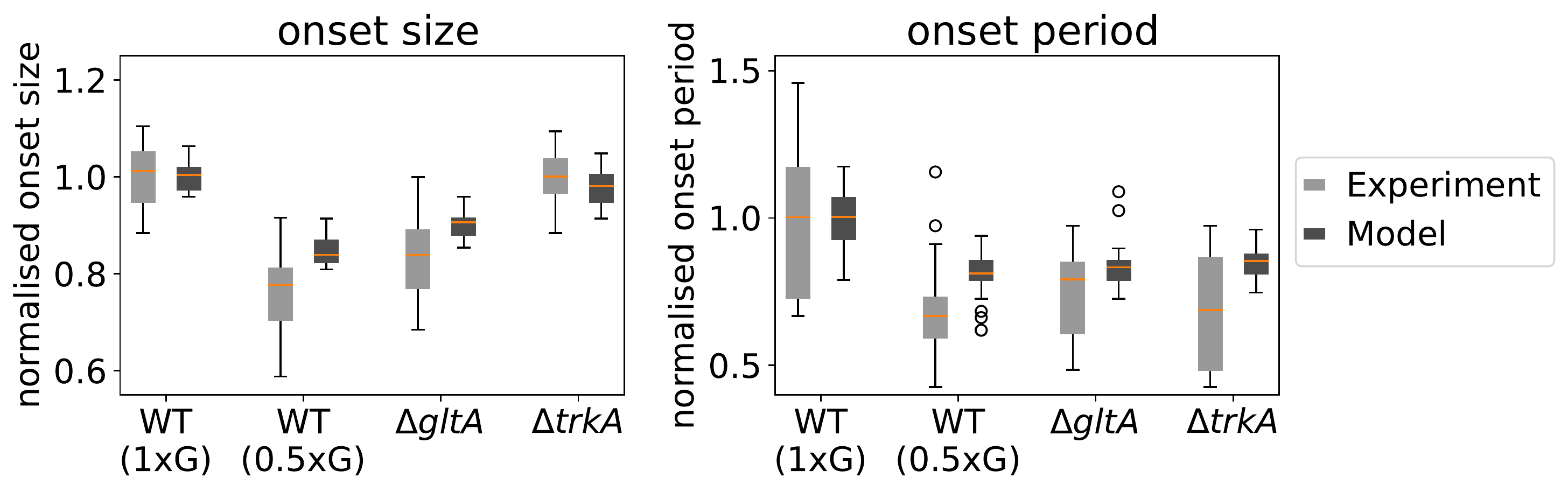}}
\caption{Biofilm size and oscillation period at onset. Light grey: Experimental 
Data.  Dark grey: Model. The data is normalised with respect to the average of 
each WT condition (model and experiments separately). Boxplots extend between 
the first and third quartiles of the normalised data, with a line (orange) at 
the median. The extreme of the whiskers denote the range of the data within 1.5x 
the interquartile range, with outliers plotted individually.  Experimental data 
contains a minimum of 12 data points per condition. See Methods for details on 
the simulations.}
\label{fig_onsets}
\end{figure}

According to the aforementioned mechanism for the oscillations, if a biofilm 
increases its glutamate consumption, or if the glutamate concentration in the 
media is reduced, oscillations should start earlier because the centre becomes 
stressed at a smaller biofilm size.
This expectation is fulfilled when simulations are performed with halved 
glutamate concentration in the media ($G_{m}$), in good agreement with the 
experimental data (compare the first two conditions in the left panel of 
Fig.~\ref{fig_onsets}, see also Refs.~\citep{Liu2015, Martinez-Corral2018}). 
Similarly, bacteria that cannot synthesise glutamate due to a deletion in the 
\emph{gltA} gene are expected to consume more glutamate from the media. We model 
this condition as an increase in both glutamate uptake rate ($\alpha_{g}$) and 
degradation rate ($\delta_g$), as a result of higher glutamate demand. Also in 
this case oscillations start earlier both in the model and the experiments 
(third condition in the left panel of Fig.~\ref{fig_onsets}).

In these two cases of reduced onset size, the onset period is also smaller than 
in the reference situation (first three conditions in the right panel of 
Fig.~\ref{fig_onsets}). As we have described in previous work 
\citep{Martinez-Corral2018}, larger biofilms tend to have higher periods, the 
explanation being that the larger a biofilm is, the more time it takes for the 
stress signal to propagate from centre to periphery, and thus the longer the 
oscillation period. 

\begin{figure}
\centerline{\includegraphics[width=0.8\textwidth]{\figsdirsecond/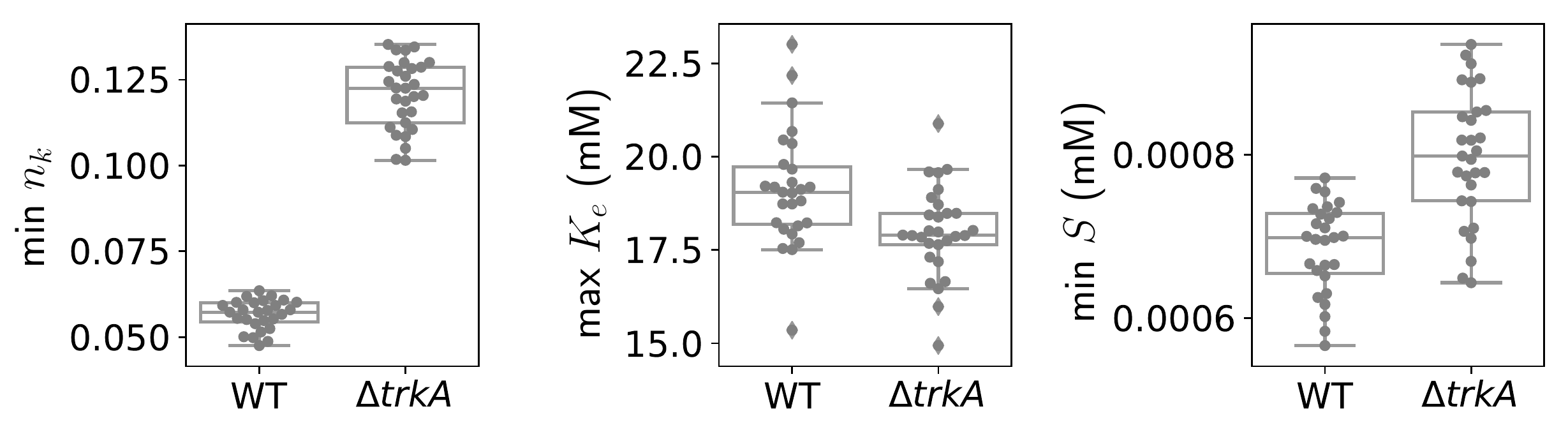}}
\caption{The $\Delta trkA$ mutation in the model leads to impaired stress 
relief. For each simulation, either the maximum or the minimum of the variable 
during the oscillations was computed for the peripheral region (outermost 100 
$\mathrm{\mu m}$). Each dot represents a simulation, from the same data as in 
Fig.~\ref{fig_onsets} from the main text.}\label{trkA_mutant}
\end{figure}

However, according to the oscillation mechanism that this model suggests, the 
period of the oscillations must also depend on the metabolism dynamics and the 
stress relieving capabilites of the cells due to the potassium effects.
The $\Delta trkA$ mutant strain was characterised in Ref. \citep{Prindle2015} to 
be deficient in electrochemical signalling, due to the deletion of the gating 
domain of the YugO channel. The mutation can be interpreted to render a channel with 
reduced conductance and increased basal leak \citep{Prindle2015}. This can be 
modelled with increased  $a_0$ (to simulate a leaky channel), and reduced $g_K$ 
(to simulate reduced conductance once the channel is open). The model 
simulations predict that oscillations start at the same size as with the basal 
parameters, but with a reduced period, in good agreement with the experimental 
data (last condition in Fig.~\ref{fig_onsets}). The unaffected onset size is 
consistent with the fact that glutamate metabolism is largely unperturbed in 
this mutant, whereas the altered potassium channel leads to less effective 
stress relief during wave propagation (Figs.~\ref{trkA_mutant}), and thus 
facilitates a subsequent cycle. Therefore, the relationship between biofilm size 
and oscillation period is not universal, but depends upon the metabolic and 
electrical signalling capabilities of the cells and thus on the genetic 
background of the strain.

\section{Discussion}

\new{The oscillations in growth rate and membrane potential exhibited by \emph{B. 
subtilis} biofilms is an instance of self-organising, emergent 
behaviour similar to the complex collective dynamics that characterizes neuronal tissue. }
Here we have proposed a unified conceptual framework for these bacterial oscillations that 
combines glutamate metabolism with potassium wave propagation.
Our discrete reaction-diffusion scheme assumes an initially homogeneous 
population of cells, and exhibits a spontaneous emergence of oscillations beyond 
a critical biofilm size.
Our work shows that oscillations are triggered by a decrease in glutamate 
levels in the biofilm centre, that lead to metabolic stress and potassium 
release.
Potassium diffusion to the neighbors interferes with glutamate metabolism, thus 
causing a self-propagating wave of membrane potential changes that allows 
glutamate recovery in the centre and thus explains the oscillation cycle. 

%We notice that there are some differences in the ThT profile between the 
%experiments and the model, which are likely the result of our highly simplified 
%treatment of the membrane potential.
%Due to the relevance of the regulation of membrane potential for multiple 
%processes in cells, there are likely to be other processes involved in its 
%regulation which we have not considered.
%In addition, potassium uptake and release is also likely to be fine-tuned by 
%other processes.
%All this is likely to be necessary to ultimately explain these oscillations with 
%a finer quantitative detail, which we expect will be addressed in future 
%studies.  

With some notable recent exceptions in multicellular eukaryotes \cite{Levin:2017fk}, electrophysiology has only been considered relevant so far for excitable cells such as those in cardiac and neural tissue.
\new{The resemblance, both at the molecular and functional level, between excitable wave propagation in these tissues and the communication of metabolic stress among bacterial cells calls for an evolutionary perspective on the phenomenon of electrical signalling.
Animal nervous systems have evolved intricate electrochemical circuits that allow sensing and responding to both internal and external stimuli, with high integrative and computational capabilities clearly evidenced by the human brain. How this complexity arose is still a matter of active investigation \cite{Moroz2009, Bucher2015,Kristan2016,Moroz2016} but it is tempting to speculate that the bacterial oscillations reported here may represent an ancient instance of electrically-mediated information transmission.}

\new{Primitive prokaryotic potassium channels are regarded as the ancestors of the animal cation channels whose diversification has been linked to the evolution of nervous systems \cite{Anderson2001,Zakon2012}.
Given that high intracellular concentrations of potassium are essential for bacterial pH maintenance \cite{Armstrong2015, Gundlach2018}, it is natural to expect that potassium channels were used in early bacteria for osmoregulation \cite{Kristan2016}.
The biofilm oscillations studied here suggest that potassium also acts as a signalling intermediate in modern bacteria, communicating information among cells of both the same and also different bacterial species \cite{Humphries2017}. }

\new{The link between metabolism and electrical signalling described here is also reminiscent of the relationship between cellular energy and neuronal function \cite{Yu2017}.
Neuronal dynamics is highly dependent upon cellular metabolism and energy state: the high energy requirements of the ion pumps that maintain proper electrochemical gradients are well known \cite{Magistretti2015}.
In addition, some animal voltage-gated ion channels directly respond to energy-related metabolites like NAD(H) and ADP/ATP through direct ligand binding \cite{Kilfoil2013}, as in the bacterial YugO channel. }

\new{While the mechanisms that allow information processing in bacterial biofilms share similar chemical and electrical processes with animal brains, the latter compute at much faster speeds and thus in much more complex ways.
This can be partly ascribed to the reliance of animal brains on faster voltage-gated ion channels, such as those dependent on sodium, which evolved only when rapid responses were increasingly beneficial (due for instance to the appearance of predators) \cite{Kristan2016}.
Nevertheless, slow propagation of electrical activity is still present in modern animal brains. During the phenomenon of cortical spreading depression, for instance, a slow wave of electrical activity propagates over brain tissue, with massive ionic redistribution  involving, critically, potassium and glutamate accompanied by a strong metabolic disturbance. This phenomenon has been regarded as intrinsic to neurons and is evolutionarily conserved across animal species \citep{Ayata2015}.
In the light of these facts, it is tempting to see the potassium-mediated transmission of stress among bacterial cells as a precursor of more recent electrically-mediated information propagation tasks in animal brains.
The study of membrane potential dynamics and electrical signalling processes in other bacterial species will be important to further illuminate this issue.
}

\section*{Methods}

\subsection{Experimental data}

\paragraph{Strains and Plasmids.}
All experiments were performed using \emph{Bacillus subtilis} strain NCIB 3610. 
For the $\Delta trkA$ mutant, we deleted the C-terminal portion of \emph{yugO} 
(amino acids 117Ð328), leaving only the N-terminal ion channel portion of YugO 
(amino acids 1Ð116). For the $\Delta gltA$ mutant, the \emph{gltA} gene was 
replaced by kanamycin resistant gene.

\paragraph{Growth conditions.}
Biofilms were grown using the standard MSgg biofilm-forming medium 
\cite{Branda:2001fk}. In this media glutamate is the only nitrogen source for 
the bacteria. We explored biofilm dynamics at glutamate concentrations of 
1$\times$ (30~mM) and 0.5$\times$ (15~mM). For microfluidics, We used the 
CellASIC ONIX Microfluidic Platform and the Y04D microfluidic plate (EMD 
Millipore). Details can be found in our previous work \cite{Liu2015,Liu:2017uq}.

\paragraph{Time-Lapse Microscopy.}
Biofilms were monitored using time-lapse microscopy, using Olympus IX81 and IX83 
inverted microscopes. To image entire biofilms, 10$\times$ lens objectives were 
used. Images were taken every 10~min. We tracked membrane potential dynamics 
using the fluorescent dyes Thioflavin T (10~$\mu$M) and APG-4 (2$\mu$M), from 
TEFLabs \cite{Prindle2015}.

\subsection{Modeling}

\paragraph{Simulation methods.}
Simulations were performed using custom-code in C, and analysis and plotting was 
done in Python. We used a first-order approximation of the discrete Laplacian 
operator \citep{abramowitz} and the time evolution of the variables was obtained 
using a 4th order Runge-Kutta method. Boundary conditions were reflective: the 
outside neighbours of the lattice sites at the boundary of the system are the 
boundary sites themselves (such that the discrete normal derivatives of the 
variables at the boundary are zero).
The simulation time step was set to $5\times10^{-6}$ h, and each lattice site 
corresponds to 10 $\mathrm{\mu m}$. 
The system was allowed to relax towards the steady state at the beginning of the 
simulation by preventing expansion during 1 hour of simulated time. 
The initial conditions for the intracellular variables are $S=0$ $\mu M$, 
$n_{k}=0$, $V=-156$ mM, $K_{i}=300$ mM, $ThT=0$ $\mu M$, $Rg=1$ $\mu M$, 
$H_{i}=0.5$, $H=0.25$, $R=1$, and $G_{i}$ is log-normally distributed around 3 
mM. In the biofilm cells, $K_{e}$ is log-normally distributed around 8 mM, and 
$A=1$ mM. In the non-biofilm cells, $A=0$. Initial $G_{e}=30$ mM (or 15 mM in reduced glutamate concentrations).

\paragraph{Numerical perturbations.}
In order to study the effect of halved glutamate in the media, and the 
\emph{trkA} and \emph{gltA} deletions, we assumed the following parameter values 
with respect to the WT: $\Delta gltA$ mutant: $\delta_g = 1.5\times$, 
$\alpha_{gt} = 1.3\times$; $\Delta trkA$ mutant: $a_0 = 2\times$, $g_K = 
0.5\times$. We performed 30 simulations per condition starting from a random 
initial radius between 150 and 250 $\mu m$. 
Potassium peaks were defined based on a threshold of 9.5 mM (pulse begins when 
crossing the threshold). Onset period is defined to be the time between the 
beginnings of the first two peaks, and onset size that of the first beginning. 

\section*{Funding}
This work was supported by the Spanish Ministry of Economy and Competitiveness 
and FEDER (project FIS2015-66503-C3-1-P), and by the Generalitat de Catalunya 
(project 2017 SGR 1054).
R.M.C. acknowledges financial support from La Caixa foundation.
J.G.O. acknowledges support from the ICREA Academia programme and from the 
``Mar\'ia de Maeztu'' Programme for Units of Excellence in R\&D (Spanish 
Ministry of Economy and Competitiveness, MDM-2014-0370). G.M.S. acknowledges 
support for this research from the San Diego Center for Systems Biology (NIH 
grant P50 GM085764), the National Institute of General Medical Sciences (grant 
R01 GM121888), the Defense Advanced Research Projects Agency (grant 
HR0011-16-2-0035), and the Howard Hughes Medical Institute-Simons Foundation 
Faculty Scholars program.
A.P. is supported by a Simons Foundation Fellowship of the Helen Hay Whitney 
Foundation and a Career Award at the Scientific Interface from the Burroughs 
Wellcome Fund.

\bibliography{extended_b5}
\clearpage

\beginsupplement
{\large{\textbf{SUPPLEMENTARY MATERIAL}}}
\maketitle

\section{Model equations and description}

As explained in the main text, we consider a one-dimensional array of simulation lattice sites where biochemical species react and diffuse. Here we describe in detail the model equations.  

\subsection{Metabolic component}
In the 'non-biofilm' sites, we consider media flow and diffusion of glutamate ($G$) and 
ammonium ($A$) according to the following equations:
\begin{align}
\frac{dG_e}{dt}&=\phi\,(G_{m}-G_e)+ D_g\nabla^2\,G_e\label{eq:Ge}\\
\frac{dA}{dt}&=\phi\,(A_{m}-A)+D_a\nabla^2\,A\label{eq:A}
\end{align}
In the microfluidics chamber, media is flowing constantly. Thus, we simulate the effect of the media flow with the first term of each equation, such that with some rate $\phi$, the concentration of the chemical species tends 
to equate that in the medium ($X_m$). The second term models diffusion.

In the biofilm, we assume that the diffusion coefficient and flow rate of 
glutamate decay exponentially with the distance to the biofilm edge $d_{e}$, due 
to the extracellular matrix and high cell density, according to the following functions:
\begin{align}
\Lambda_{\phi}&=\frac{\exp(-\gamma_{\phi}\,d_{e})+a_{\phi}}{1+a_{\phi}}\\
\Lambda_{D}&=\frac{\exp(-\gamma_{D}\,d_{e})+a_{D}}{1+a_{D}}
\end{align}
These functional forms are chosen such that at the edge $\Lambda_x=1$, and flow 
and diffusion match those in the media. The coefficients decay towards the 
biofilm interior, tending asymptotically to $a/(1+a)$ in the centre, which 
ensures some remaining flow and diffusion. We follow the original assumption \cite{Liu2015}
that ammonium diffusion over the biofilm is very fast, and do not apply these 
reduction terms to this variable.

The dynamics of extracellular ($G_e$) and intracellular 
($G_i$) glutamate in the biofilm are modelled with the following dynamical equations:
\begin{align}
\frac{dG_e}{dt}&=-\alpha_{g}\,{\cal 
F}(V)\,R_{g}\,\frac{G_e}{k_{g}+G_e}+\Lambda_{\phi}\,\phi\,(G_{m}-G_e)+\Lambda_{D
}\, D_g\nabla^2\,G_e \label{eq_Ge}\\
\frac{dG_i}{dt}&=\alpha_{g}\,{\cal 
F}(V)\,R_{g}\,\frac{G_e}{k_{g}+G_e}-\alpha_a\,H\,G_i-\delta_g\,G_i\,r\,, 
\label{eq_Gi}
\end{align}
where, as mentioned, $G_e$ is affected by media flow and diffusion. The 
first term in the right-hand side of the two equations represents glutamate 
transport into the cells. As explained in the main text, we consider that glutamate transport into the cell is modulated by the 
membrane potential $V$, such that depolarisation reduces entry, and 
hyperpolarisation enhances it, according to the functional form given in Eq.~(\ref{eq_Ginip}) of the main text.

In addition, we assume that glutamate is imported into the cells through the glutamate transporter $R_{g}$, 
which saturates for large enough $G_e$, with half-maximum 
concentration $k_{g}$. We describe explicitly the dynamics of $R_{g}$ by:
\begin{align}
\frac{dR_{g}}{dt}&=\alpha_{R}-\delta_{R}\,R_g+\beta_{R}\frac{r^{n_R}}{k_{R}^{n_R
}+r^{n_R}}\,.\label{eq_Rg}
\end{align}
We thus assume that $R_{g}$ is produced at a basal rate $\alpha_{R}$ and 
degraded at a rate $\delta_{R}$. The last term accounts for the higher glutamate uptake by metabolically active cells, such that the presence of biomass-producing 
biomolecules, such as ribosomal proteins, denoted by $r$, enhances $R_g$ synthesis via a Hill function 
with exponent $n_R$.

Equation~(\ref{eq_Gi}) also assumes that intracellular glutamate concentration 
decays due to ammonium production via the GDH enzyme (represented by $H$ in the 
$\alpha_a$-term at the right-hand side of the equation) and through various 
metabolic tasks including in particular biomass production ($\delta_g$-term). 

The production of ammonium is regulated by the activity of the 
enzyme GDH. We describe the dynamics of the inactive and active forms of this 
enzyme, $h$ and $H$ respectively, by the equations:
\begin{align}
\frac{dh}{dt}&=\frac{\alpha_{h}}{1+(G_i/k_{h})^{n_{h}}}  
-\alpha_H\,h\,\frac{G_i^{n_H}}{k_{H}^{n_{H}}+G_i^{n_{H}}}-\gamma_{h}\,
h+\gamma_H\,H\label{eq_h}\\
\frac{dH}{dt}&=\alpha_H\,h\,\frac{G_i^{n_H}}{k_{H}^{n_{H}}+G_i^{n_{H}}}
-\gamma_H\,H\label{eq_H}
\end{align}
such that we account for synthesis and degradation of inactive GDH and its conversion 
into the active form ($\alpha_H$-term in the two equations).
As explained in the main text, we assume that high concentrations of glutamate inhibit GDH synthesis, whereas 
activation is positively regulated by glutamate via a Hill function with 
exponent $n_H$.
We also consider deactivation at a constant rate $\gamma_H$.

Ammonium dynamics is affected by production from glutamate by active GDH, 
consumption for various metabolic processes such as biomass production, and 
diffusion:
\begin{align}
\frac{dA}{dt}&=\alpha_a\,H\,G_i-\delta_a\,A\,r+D_a\nabla^2\,A\label{eq_A}
\end{align}
Finally, biomass production is considered to increase with ammonium and 
intracellular glutamate, and to be subject to linear decay:
\begin{align}
\frac{dr}{dt}&=\beta_r\,A\,G_i-\gamma_r\,r\label{eq_r}
\end{align}

\subsection{Electrical signalling}

Next we incorporate an adapted version of the electrical model introduced in \citep{Prindle2015}. As explained in the main text, we consider an inhibitory effect of intracellular glutamate on a stress variable $S$, whose production rate is modelled 
with an inhibitory Hill function: 
\begin{align}
\frac{dS}{dt}&=\frac{S_0}{1+\bigl(\frac{G_i}{G_{s0}}\bigr)^{n_s}}-\gamma_s\,
S\label{eq_SA}
\end{align}

We explicitly consider both extracellular ($K_e$) and intracellular ($K_i$) 
potassium, whose dynamics are given by:
\begin{align}
\frac{dK_e}{dt}&=F\, g_K\,n_k^4\,(V-V_{K})- D_p\,G_i\,K_e(K_{i0}-K_i)+
\Lambda_{\phi}\,\phi\,(K_{m}-K_e)+\Lambda_{D}D_k\nabla^2\,K_e\label{eq_Ke}\\
\frac{dK_i}{dt}&=-F\,g_K\,n_k^4\,(V-V_{K})+ 
D_p\,G_i\,K_e(K_{i0}-K_i)\label{eq:Ki}
\end{align}

Potassium uptake is assumed to be governed by homeostatic processes that tend to 
keep its intracellular concentration at a fixed value $K_{i0}$, described by the 
second term in the right-hand side of Eqs.~(\ref{eq_Ke})-(\ref{eq:Ki}).
Uptake is also made to depend on the metabolic state (glutamate level), to 
account for the energy demand of the process. In addition, extracellular 
potassium diffuses and is subject to the media flow in the chamber.

Potassium flow through its ion channel (first term in the right-hand side of the 
$K_e$ and $K_i$ equations) is governed by the corresponding Nernst potential:
\begin{align}
V_{K}&=V_{K0}\,\ln\biggl(\frac{K_{e}}{K_{i}}\biggr)\,,
\end{align}
and depends on the opening probability of the potassium channel, $n_{k}$:
\begin{align}
\frac{dn_k}{dt}&=\frac{a_0\,S}{S_{th}+S}\,(1-n_k)-b\,n_k
\end{align}
In the media lattice sites, as in the case of glutamate and ammonium 
[Eqs.~(\ref{eq:Ge}) and (\ref{eq:A})], the extracellular potassium dynamics is 
affected by diffusion and by the media flow:
\begin{align}
\frac{dK_e}{dt}&=\phi\,(K_{m}-K_e)+D_k\nabla^2\,K_e\label{eq:Ke}
\end{align}
The membrane potential dynamics is described by a Hodgkin-Huxley-like 
conductance-based model containing potassium flux through the ion channel and a 
leak current \citep{Prindle2015}: 
\begin{align}
\frac{dV}{dt}&=-g_K\,n_k^4\,(V-V_{K})-g_L\,(V-V_{L})\,,
\label{HHeq}
\end{align}
where the leak potential $V_{L}$ is assumed to depend on the extracellular 
potassium \cite{Prindle2015} in a threshold-linear manner, such that when $K_e$ 
is larger than its basal level in the medium, $K_m$, the leak potential $V_{L}$ 
grows linearly (and the cell depolarizes), while when $K_e<K_m$ the leak potential 
stays at its basal level:
\begin{align}
V_{L}&=V_{L0}+d_L\frac{K_{e}-K_{m}}{1-e^{-(K_{e}-K_{m})/\sigma}}
\end{align}
Finally, we include the ThT reporter $\mathcal{T}$ downstream of the membrane 
potential, increasing when the cells become hyperpolarised due to potassium 
release:
\begin{align}
\frac{d\mathcal{T}}{dt}&=\frac{\alpha_\mathcal{T}}{1+\exp(g_{\mathcal{T}}(V-V_{
0\mathcal{T}}))}-\gamma_\mathcal{T}\,\mathcal{T}
\label{eq_T}
\end{align}

\subsection{Simplified model}

In the simplified model, we keep the same dynamics for the electrical part (Eqs.~(\ref{eq_SA})-(\ref{eq_T})). The metabolic part is simplified as follows: Eqs.~(\ref{eq_h})-(\ref{eq_r}) are removed, the equations for extracellular glutamate dynamics, in both `biofilm' and `non-biofilm' sites (Eqs.~\ref{eq:Ge}, \ref{eq_Ge}), are maintained, and Eqs.~(\ref{eq_Gi}) and (\ref{eq_Rg}) become:
\begin{align}
\frac{dG_i}{dt}&=\alpha_{g}\,{\cal 
F}(V)\,R_{g}\,\frac{G_e}{k_{g}+G_e}-\delta_g\,G_i\\
\frac{dR_{g}}{dt}&=\alpha_{R}-\delta_{R}\,R_g+\alpha_{r}\frac{G_i^{n_r}}{k_{r}^{
n_r}+G_i^{n_r}}
\end{align}

\section{Supplementary tables and figures}

{\rowcolors{2}{gray!15}{white}
{\small
\begin{longtable}{p{1.7cm}|p{7cm}|p{1.8cm}|p{1.8cm}|p{2.5cm}}
\hiderowcolors
\hline
%\multirow{2}{*}{Parameter} & \multirow{2}{*}{Full model} & 
%\multicolumn{2}{c}{Simplified model} &  \multirow{2}{*}{Units}\\\cline{3-4}
%							&				
%		& \multicolumn{1}{c}{(1D)}& \multicolumn{1}{c}{(2D)} & \\
{Parameter} & {Description} & {Full model} & {Simplified model} & {Units}\\			
\hline
\endfirsthead
\multicolumn{5}{r}
{\tablename\ \thetable\ -- \textit{Continued from previous page}} \\
\hline
{Parameter} & {Description} &{Full model} & {Simplified model} &  {Units}\\\cline{3-4}
\hline
\endhead
\hline \multicolumn{5}{r}{\tablename\ \thetable\ -- \textit{Continued on next 
page}} \\
\endfoot
\hline
\caption{Parameter description and basal values for the two models.}{}\label{table_extended}\\
\endlastfoot
\showrowcolors
$\alpha_{g}$ & glutamate uptake constant & 36.0 & 24.0 & mM /($\mathrm{\mu}$M h) \\
$k_{g}$ & extracellular glutamate concentration at half-maximal uptake rate & 0.75 & 0.75 & mM \\
$G_m$ & glutamate concentration in the media & 30.0 & 30.0 & mM \\
$D_g$ & glutamate diffusion coefficient & 4e+06 & 4e+06 & $\mathrm{\mu m^2/h}$ \\
$\alpha_a$ & ammonium production constant & 4.5 & - & $\mathrm{\mu}$M$^{-1}$h$^{-1}$ \\
$\delta_g$ & glutamate degradation constant & 0.525 & 4.8 & mM$^{-1}$h$^{-1}$ \\
$\alpha_{R}$ & glutamate receptor synthesis rate & 6.75 & 4.5 & $\mathrm{\mu}$M/h \\
$\delta_{R}$ & glutamate receptor decay constant & 36.0 & 24.0 & h$^{-1}$ \\
$\beta_{R}$ & maximum rate of glutamate receptor induction & 45.0 & 31.0 & $\mathrm{\mu}$M/h \\
$k_{R}$ & threshold for $R_g$ induction & 5.0 & 2.25 & mM \\
$n_{R}$ & Hill coefficient for $R_g$ induction & 2.0 & 2.0 & - \\
$\alpha_{h}$ & maximal GDH synthesis rate & 0.075 & - & $\mathrm{\mu}$M/h \\
$k_{h}$ & threshold for inhibition of GDH synthesis by glutamate & 1.5 & - & mM \\
$\gamma_{h}$ & GDH decay rate & 0.01 & - & $\mathrm{h^{-1}}$ \\
$n_{h}$ & Hill coefficient for GDH synthesis inhibition by glutamate & 2.0 & - & - \\
$\alpha_H$ & GDH activation constant & 3.0 & - & h$^{-1}$ \\
$k_{H}$ & intracellular glutamate concentration for half-maximal GDH activation & 0.4 & - & mM \\
$n_{H}$ & Hill coefficient for GDH activation by glutamate & 2.0 & - & - \\
$\gamma_H$ & GDH deactivation constant & 5.0 & - & $h^{-1}$ \\
$\delta_a$ & ammonium consumption constant & 0.135 & - & mM$^{-1}$h$^{-1}$ \\
$A_m$ & ammonium concentration in the media & 0.0 & - & mM \\
$D_a$ & ammonium diffusion coefficient & 7e+06 & - & $\mathrm{\mu m^2/h}$ \\
$\beta_r$ & biomass-producing biomolecules synthesis rate & 15.0 & - & mM$^{-1}$h$^{-1}$ \\
$\gamma_r$ & biomass-producing biomolecules decay rate & 6.0 & - & h$^{-1}$ \\
$S_0$ & maximum stress production rate & 1.12 & 1.12 & $\mathrm{\mu}$M/h \\
$G_{S0}$ & threshold for stress inhibition by glutamate & 0.2 & 0.2 & mM \\
$n_s$ & Hill coefficient for stress inhibition by glutamate & 2.0 & 2.0 & - \\
$\gamma_s$ & stress decay constant & 2.8 & 2.8 & h$^{-1}$ \\
$a_0$ & maximum rate of increase of the potassium channel gating probability & 91.0 & 91.0 & h$^{-1}$ \\
$S_{th}$ & stress level for half-maximal gating activity & 0.03 & 0.03 & $\mathrm{\mu}$M \\
$b$ & opening probability decay constant & 21.25 & 34.0 & h$^{-1}$ \\
$F$ & membrane capacitance & 0.05 & 0.05 & mM/mV \\
$g_K$ & potassium channel conductance & 70.0 & 70.0 & h$^{-1}$ \\
$V_{K0}$ & Nernst potential prefactor & 100.0 & 100.0 & mV \\
$D_p$ & potassium uptake constant & 0.12 & 0.12 & mM$^{-2}$h$^{-1}$ \\
$K_{i0}$ & intracellular potassium concentration & 300.0 & - & mM \\
$D_k$ & potassium diffusion coefficient & 7e+06 & 7e+06 & $\mathrm{\mu m^2/h}$ \\
$K_m$ & potassium concentration in the media & 8.0 & 8.0 & mM \\
$g_L$ & leak conductance & 18.0 & 18.0 & h$^{-1}$ \\
$d_L$ & leak slope coefficient & 4.0 & 4.0 & mV/mM \\
$V_{L0}$ & basal leak potential & -156.0 & -156.0 & mV \\
$\sigma$ & leak threshold sharpness coefficient & 0.1 & 0.1 & mM \\
$g_v$ & inverse sensitivity of glutamate uptake to membrane potential & 1.0 & 1.0 & mV$^{-1}$ \\
$V_0$ & resting membrane potential & -150.0 & -150.0 & mV \\
$\alpha_\mathcal{T}$ & maximal rate of ThT uptake & 20.0 & 20.0 & $\mathrm{\mu}$M/h \\
$\gamma_\mathcal{T}$ & intracellular ThT decay constant & 10.0 & 10.0 & $\mathrm{h^{-1}}$ \\
$g_\mathcal{T}$ & inverse sensitivity of ThT to membrane potential & 0.3 & 0.3 & mV$^{-1}$ \\
$P_{grow}$ & growth probability & 0.3 & 0.5 & h$^{-1}$ \\
$\gamma_{\phi}$ & spatial decay rate of the flow rate within the biofilm & 0.0085 & 0.0085 & $\mathrm{\mu m}$ \\
$a_{\phi}$ & basal flow factor & 0.012 & 0.012 & - \\
$\gamma_{D}$ & spatial decay rate of the diffusion coefficient within the biofilm & 0.0085 & 0.0085 & $\mathrm{\mu m}$ \\
$a_{D}$ & basal diffusion factor & 0.012 & - \\
$\phi$ & flow rate & 5.0 & 5.0 & h$^{-1}$ \\
\end{longtable}}}

\clearpage

\end{document}